\documentclass[twoside,english,6pt, sort&compress]{elsarticle}
\usepackage{lmodern}

\usepackage[T1]{fontenc}
\usepackage[latin9]{inputenc}
\usepackage{geometry}
\usepackage{array}
\usepackage{amstext}
\usepackage{graphicx}
\usepackage{esint}
\usepackage{subscript}

\makeatletter

\providecommand{\tabularnewline}{\\}
\usepackage{caption}

\journal{-}

\makeatother

\usepackage{babel}
\begin{document}

\begin{frontmatter}{}

\title{Steps towards the experimental realization of surface plasmon polariton
enhanced spontaneous parametric down-conversion}

\author[rvt]{A. Loot\corref{cor1}}

\ead{ardi.loot@ut.ee}

\author[rvt]{I. Sildos}

\author[rvt]{V. Kiisk}

\author[focal]{T. Romann}

\author[rvt]{V. Hizhnyakov}

\cortext[cor1]{Corresponding author}

\address[rvt]{Institute of Physics, University of Tartu, W. Ostwald St. 1, 50411
Tartu, Estonia}

\address[focal]{Institute of Chemistry, University of Tartu, Ravila 14a, 50411 Tartu,
Estonia}
\begin{abstract}
The realization of efficient and miniature source of entangled
photon pairs stills remains a challenge. In this work, we experimentally
studied the possibility to enhance the process of spontaneous parametric
down-conversion by surface plasmon polaritons. The details of the
experimental setup will be discussed along with theoretical calculations
of the enhancement factor.
\end{abstract}

\end{frontmatter}{}

\section{Introduction}

Efficient source of entangled photon pairs is essential for many quantum
optical experiments and applications like quantum communication, quantum-key
distribution, quantum computing, etc \citep{Mandel1995,Gerry2004}.
Up to now, the most perspective source of entangled photon pairs is
based on spontaneous parametric down-conversion (SPDC) \textendash{}
splitting of pump photons into signal and idler photons due to nonlinear
interaction with matter. Most commonly, the SPDC takes place in a
bulk crystal, however, the practical availability is limited by the
low efficiency of SPDC (only around $10^{-12}$) and large footprint
of the crystal (several millimeters) \citep{Burnham1970,Klyshko1967,Klyshko1988}.
To overcome the problem of the low efficiency, the process of SPDC
have been studied in nonlinear periodically-poled waveguides by many
research groups and efficiencies up to $10^{-6}$ have been reported
\citep{Tanzilli2001,Sanaka2001a,Banaszek2001,Mason2002,Bock2016}.
However, the problem of large footprint still remains as a limitation
of practical usability in miniature devices.

An idea to use the field enhancement of surface plasmon polaritons
(SPPs) to boost the efficiency of SPDC was theoretically studied in
Refs. \citep{Loot2015,Loot2018} and enhancements, also accessible
in miniature sources, up to $40\cdot10^{3}$ were predicted. For the
proof of principle, the SPP-enhanced SPDC were studied in the structure
similar to Kretschmann configuration (see Fig. \ref{fig:spp-spdc-struct}),
so splitting pump photons into two SPP modes (denoted by $\beta_{s}$
and $\beta_{i}$ in Fig. \ref{fig:spp-spdc-struct}) instead to photonic
modes becomes possible if phase-matching conditions (see Ref. \citep{Loot2015})
\begin{eqnarray}
\frac{1}{\lambda_{p}} & = & \frac{1}{\lambda_{s}}+\frac{1}{\lambda_{i}},\\
k_{p}\left(\lambda_{p},\theta_{p}\right) & = & \beta_{s}\left(\lambda_{s}\right)+\beta_{i}\left(\lambda_{i}\right),\label{eq:pm-conds}
\end{eqnarray}

\noindent are fulfilled, where $\lambda_{p},$ $\lambda_{s}$ and
$\lambda_{i}$ denote the wavelength of the pump, signal and idler,
respectively, $k_{p}=2\pi n_{p}\sin\theta_{p}/\lambda_{p}$ is the
tangential wavevector component of the pump beam, $\theta_{p}$ is
the angle of incidence of the pump beam inside the prism, $n_{p}$
is the refractive index of the prism, $\beta_{s}$ and $\beta_{i}$
are the propagation constants of the signal and idler SPP modes, respectively.
The enhancement of SPDC follows from the field enhancements of two
participating plasmonic modes and the total enhancement of SPDC is
given by

\begin{equation}
\Upsilon=\eta_{p}^{2}\eta_{s}^{2}\eta_{i}^{2},\label{eq:spdc-enh}
\end{equation}

\noindent where $\eta_{p},$ $\eta_{s}$ and $\eta_{i}$ denote the
field enhancement factors of the pump, signal and idler, respectively
\citep{Loot2015}. After propagation, the signal and idler plasmons
leak back to the prism as photons.

The goal of this work is the experimental realization of SPP-enhanced
SPDC in the structure shown in Fig. \ref{fig:spp-spdc-struct}. To
do that, goniometric measurement setup with lock-in amplification
was designed. To the best of our knowledge, it is the first experimental
attempt to plasmonically enhance SPDC.

\section{Experimental setup}

\subsection{Structure}
\noindent \begin{flushleft}
\begin{minipage}[t]{7.75cm}%
\noindent \begin{center}
\includegraphics[width=4cm]{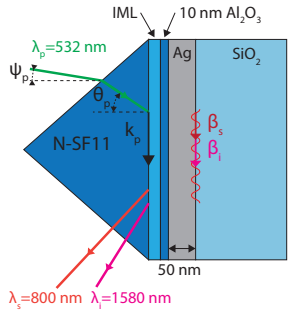}
\par\end{center}
\captionof{figure}{The structure for SPP-enhanced SPDC: N-SF11 prism, index-matching liquid (IML), Al\textsubscript{2}O\textsubscript{3} layer,  silver layer and single crystal alpha quartz. The angle of incidence of the pump laser ($\lambda_{p}=532\,\text{nm}$) is $\Psi_p$ outside the prism and $\theta_p$ inside the prism. The propagation constants of signal ($\lambda_{s}=800\,\text{nm}$) and idler ($\lambda_{i}=1580\,\text{nm}$) plasmons are $\beta_s$ and  $\beta_i$ respectively.}

\label{fig:spp-spdc-struct}%
\end{minipage}\hspace{0.5cm}%
\begin{minipage}[t]{7.5cm}%
\captionof{table}{The parameters of the structure in Fig. \ref{fig:spp-spdc-struct} extracted by fitting the reflection curves in Fig. \ref{fig:spp-resonances-shg}a.}

\setlength{\extrarowheight}{4pt}

\begin{small}
\noindent \begin{center}
\begin{tabular}{|>{\centering}p{1.2cm}||>{\centering}m{1cm}|>{\centering}m{1.75cm}||>{\centering}m{1cm}|>{\centering}m{1.75cm}|}
\hline 
 & \multicolumn{2}{c||}{$\lambda=802\,\text{nm}$} & \multicolumn{2}{c|}{$\lambda=1064\,\text{nm}$}\tabularnewline
\hline 
\hline 
 & $d$~(nm) & $n$ & $d$~(nm) & $n$\tabularnewline
\hline 
N-SF11 &  & $1.765$ &  & $1.754$\tabularnewline
\hline 
Al\textsubscript{2}O\textsubscript{3} & $10.0$ & $1.76+0.055i$ & $10.0$ & $1.76+0.055i$\tabularnewline
\hline 
Ag & $52.9$ & $0.146+5.80i$ & $50.8$ & $0.204+8.07i$\tabularnewline
\hline 
SiO\textsubscript{2} &  & $1.538$ &  & $1.534$\tabularnewline
\hline 
\end{tabular}
\par\end{center}
\label{tab:structure-fit-params}

\end{small}%
\end{minipage}
\par\end{flushleft}

The structure for the realization of SPP-enhanced SPDC is shown in
Fig. \ref{fig:spp-spdc-struct}. First, polished X-cut single crystal
quartz substrate ($10\times10\times0.5\,\text{mm}^{3}$) was bought
from PI-KEM Ltd. It acts as a nonlinear crystal (second-order susceptibility
$\chi_{\text{xxx}}^{(2)}=0.6\,\text{pm/V}$) and is selected due to
relatively small refractive index ($\approx1.55$) in comparison to
the other commercially available nonlinear crystals \citep{Dmitriev1999}.
Next, approximately $50\,\text{nm}$ thick silver film was deposited
on one side of the quartz crystal and subsequently covered with approximately
$10\,\text{nm}$ thick aluminum oxide layer in order to protect the
silver surface. The deposition was done by AJA International ultra-high
vacuum magnetron sputtering system at room temperature. The silver
film was deposited using $2\,\text{inch}$ target, $3\,\text{mTorr}$
argon pressure, $55\,\text{W}$ direct current power, $10\,\text{W}$
biasing RF (radio frequency) power at the sample (to make the silver
surface smoother) and the deposition time was $300\,\text{s}$. The
aluminum oxide was deposited using $3\,\text{inch}$ target, $3\,\text{mTorr}$
argon pressure, $150\,\text{W}$ RF source and deposition time $250\,\text{s}$.
Shortly before measurements, the silver-quartz structure was attached
to the high refractive index N-SF11 right-angle prism (Edmund Optics
\#47-276, side length $15\,\text{mm}$) with index-matching liquid
(IML) from Cargille (Series M, refractive index $1.77$).

\subsection{Goniometric setup}

\begin{figure}
\noindent \centering{}%
\begin{tabular}{>{\raggedright}m{7.5cm}>{\raggedright}m{8.5cm}}
\includegraphics[width=7.5cm]{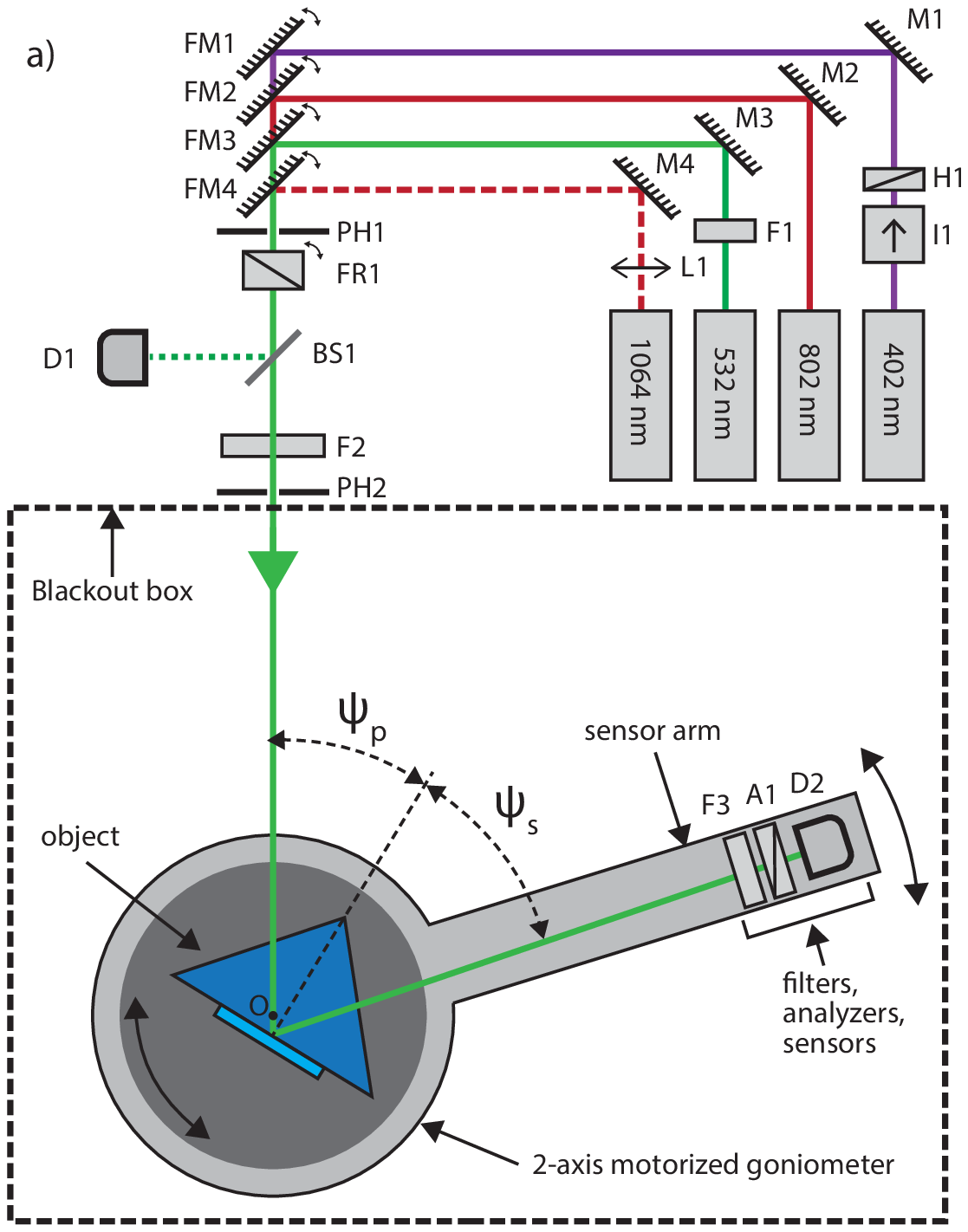} & \includegraphics[width=8.5cm]{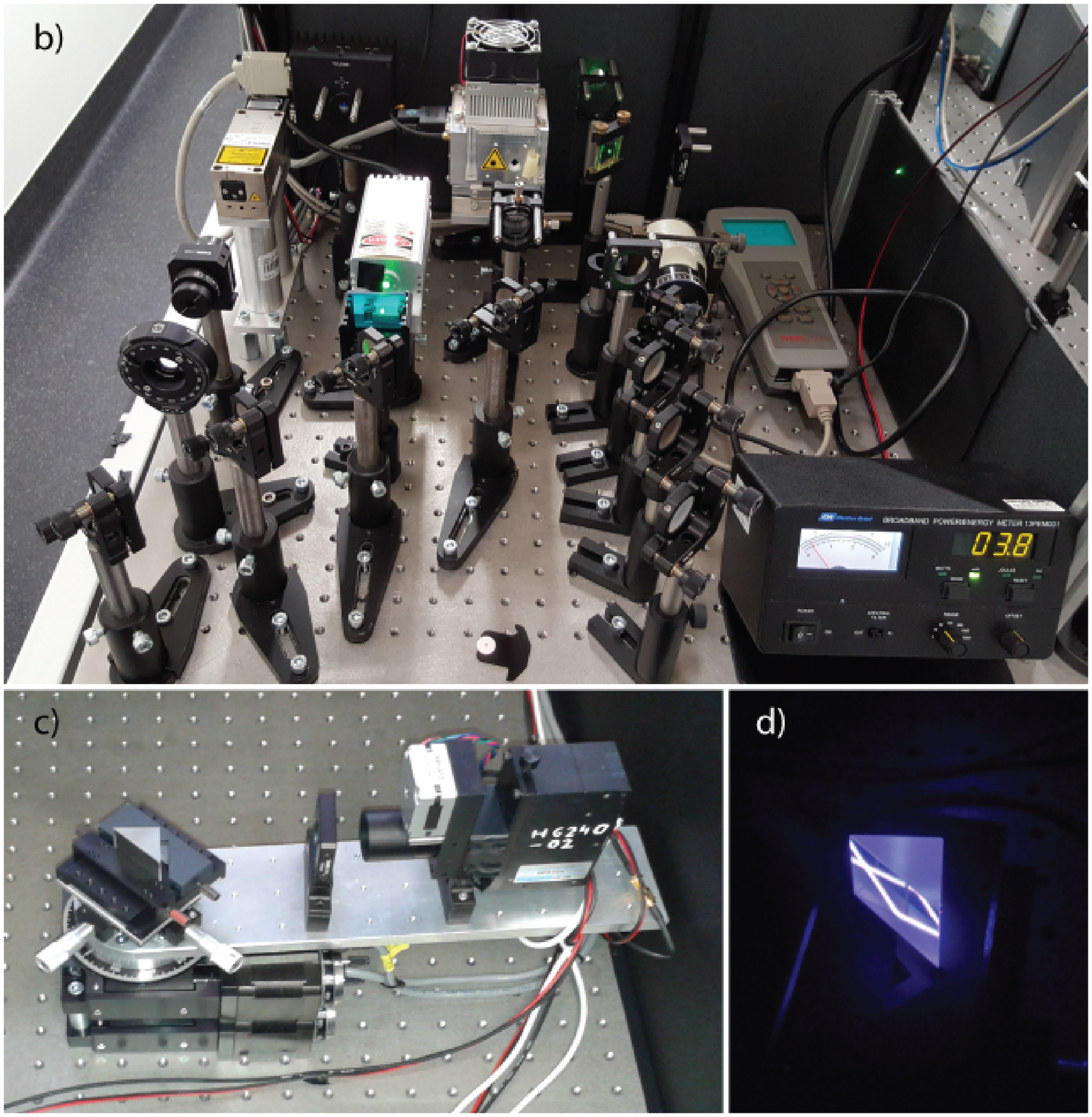}\tabularnewline
\end{tabular}\caption{The schematic representation of the experimental setup (a) and the
pictures of the experimental setup: lasers (b), goniometer with SPDC
detection setup (c) and N-SF11 prism under illumination of $402\,\text{nm}$
laser light (d). \label{fig:experimental-setup-struct}}
\end{figure}

The structure in Fig. \ref{fig:spp-spdc-struct} was measured in custom-made
two-axis motorized goniometer schematically shown in Fig. \ref{fig:experimental-setup-struct}.
The goniometer consists of two computer-controllable rotational stages
from STANDA (8MR190V-2-VSS42), which are mounted on top of each other
to independently rotate the object and the sensor around it \citep{Loot2013}.
Lasers are stationary on the optical table and the stepper motors
of the rotational stages are driven by a two-axis controller (STANDA
8SMC1-USBhF-B2-MC2) allowing independent control of the angle of incidence
($\Psi_{p}$) and sensor position ($\Psi_{s}$) with a resolution
up to $0.0013^{\circ}.$

The structure under the study could be illuminated by four different
laser wavelengths: $402\,\text{nm}$ (Toptica iBeam smart $120\,\text{mW}$),
$802\,\text{nm}$ (Thorlabs LD808-SA60 $60\,\text{mW}$ mounted into
Thorlabs TCLDM9 and controlled by Thorlabs LDC205C), $532\,\text{nm}$
(Sintec ST-532-ST-I $300\,\text{mW}$) and $1064\,\text{nm}$ (Horus
HLX-I E020-12101 $30\,\text{mW,}$ pulsed $20\,\text{kHz}$ $1\,\text{ns}$).
The violet laser ($402\,\text{nm})$ is protected by optical isolator
I1 (Thorlabs IO-5-405-LP) and the linearly polarized output is rotated
by half-wave plate H1 (Thorlabs WPH05M-405). The green laser ($532\,\text{nm}$)
is additionally spectrally cleaned by color filter F1 and the divergent
output of the $1064\,\text{nm}$ laser is collimated by lens L1. All
lasers are linearly polarized.

The beams of the lasers are aligned to be parallel to the surface
of the optical table and to coincide the axis of the goniometer (denoted
by O) with a pair of mirrors (Thorlabs BB1-E02 for $402\,\text{nm}$
and Thorlabs PF10-03-P01 for others), where FM1 - FM4 are flip mounted
for easy switching between the wavelengths. Two pinholes, PH1 and
PH2, are installed to simplify the alignment process. The polarization
angle of lasers (except $402\,\text{nm}$) is adjusted by flip mounted
Fresnel rhomb FR1. A small fraction of the laser light is reflected
to the photodiode D1 (Thorlabs PM100 with head S130A) by the quartz
beamsplitter BS1 to monitor the laser power during measurements. Finally,
before the sample, the laser light is spectrally cleaned by filters
F2 that will be shortly specified.

On the detection side, several different configurations are possible
\textendash{} it is possible to use different filters F3, motorized
polarization analyzer A1 and different detectors D2. The concrete
detection configuration will be specified shortly for the measurements
of reflection curves, second-harmonic generation (SHG) and SPDC.

For the angular calibration of the goniometer and for the positioning
of the object, the procedures given in Ref. \citep{Loot2013} was
followed.

\subsection{Measurement of reflection curves}

The SPP resonances of the structure in Fig. \ref{fig:spp-spdc-struct}
were measured at two wavelengths ($802\,\text{nm}$ and $1064\,\text{nm}$).
To do that, lasers were p-polarized with the Fresnel rhomb FR1, the
cleanup filter F2 was not necessary and on the sensor arm only the
detector D2 was used. In the case of wavelength $802\,\text{nm},$
photodiode Thorlabs PM100 with head S130A was used and in the case
of the pulsed $1064\,\text{nm}$ laser, Melles Griot 13 PEM 001 with
thermopile sensor head was used.

\subsection{Measurement of SHG}

For the measurement of SPP-enhanced SHG, the structure was excited
by the p-polarized pulsed $1064\,\text{nm}$ laser, which was spectrally
cleaned by three filters (Semrock FF01-593-LP, BLP01-633R and FF01-715-LP)
in the position F2. On the detection side, the laser light was filtered
out by Semrock FF01-535/150-25 and FF03-525/50-25 in the position
F3 and the signal was detected through the p-polarized analyzer (A3)
by Hamamatsu H6240-02 photo-multiplier tube (PMT) in the position
D2. The TTL pulses from the PMT were counted by LabJack U6 in counter
mode with collection time $1\,\text{s}$ and the distance of PMT from
the rotational axis O was $22.3\,\text{cm}.$

\subsection{Measurement of SPDC\label{subsec:Detection-of-SPDC}}

In the case of the measurement of the SPP-enhanced SPDC (shown in
Fig. \ref{fig:spp-spdc-struct}), two filters (Semrock FF01-535/150-25
and FF03-525/50-25) were added to the experimental scheme (F2) to
clean the $532\,\text{nm}$ pump laser.

On the detection side, three interference filters (one Semrock BLP01-633R
and two Semrock FF01-800-12) were used to filter out the pump laser
(F3). As a detector (D2), Hamamatsu H6240-02 PMT was used (sensitive
up to $950\,\text{nm}$, quantum efficiency around $\approx2\,\%$
at $\lambda=800\,\text{nm}$). The distance of PMT from the rotational
axis O was $19.5\,\text{cm}$ and the resulting angular detection
span $2.3^{\circ}.$

To boost the signal-to-noise ratio of the measurements, digital lock-in
amplifier was used by placing a rotating polarization analyzer (Thorlabs
LPVIS050) to the position A1. The analyzer was rotated by a custom-built
setup consisting of a stepper motor (Phidgets NEMA-17 20mm) and its
controller (PhidgetStepper Bipolar HC 1067). The rotation frequency
of the analyzer was $f_{r}=2\,\text{Hz},$ which corresponds to the
modulation frequency $f_{m}=2f_{r}=4\,\text{Hz}$.

As SPPs are always p-polarized, then also the outcoupled SPDC signal
is p-polarized, opposite to noise, which is unpolarized. The count
rate of the PMT is then described by Malus's Law

\begin{equation}
\Gamma\left(t\right)=\Gamma_{bg}+\Gamma_{s}\cos^{2}\left(2\pi f_{r}t\right)=\left(\Gamma_{bg}+\frac{1}{2}\Gamma_{s}\right)+\frac{1}{2}\Gamma_{s}\cos\left(2\pi f_{m}t\right),
\end{equation}

\noindent where $\Gamma_{bg}$ is the rate of background (unpolarized)
counts, $\Gamma_{s}$ is the rate of the polarized SPDC signal and
$t$ is time. The time-dependent signal $\Gamma\left(t\right)$ was
registered by LabJack U6 in a stream mode at the detection rate $40\,\text{Hz}$
over $T=60\,\text{s}$ time period. The resulting time series were
analyzed by software lock-in amplifier to extract the signal rate
$\Gamma_{s}=2\sqrt{X^{2}+Y^{2}},$ where
\begin{eqnarray}
X & = & \frac{1}{T}\int_{0}^{T}dt\,\Gamma\left(t\right)\cos\left(2\pi f_{m}t\right),\\
Y & = & \frac{1}{T}\int_{0}^{T}dt\,\Gamma\left(t\right)\sin\left(2\pi f_{m}t\right).
\end{eqnarray}
Such method was shown to produce shot noise limited results \citep{Braun2002,Clarkson2010}.

\section{Results}

\subsection{Reflection curves}

\begin{figure}
\noindent \begin{centering}
\begin{tabular}{>{\raggedright}m{8cm}>{\raggedright}m{8cm}}
a) & b)\tabularnewline
\includegraphics[width=8cm]{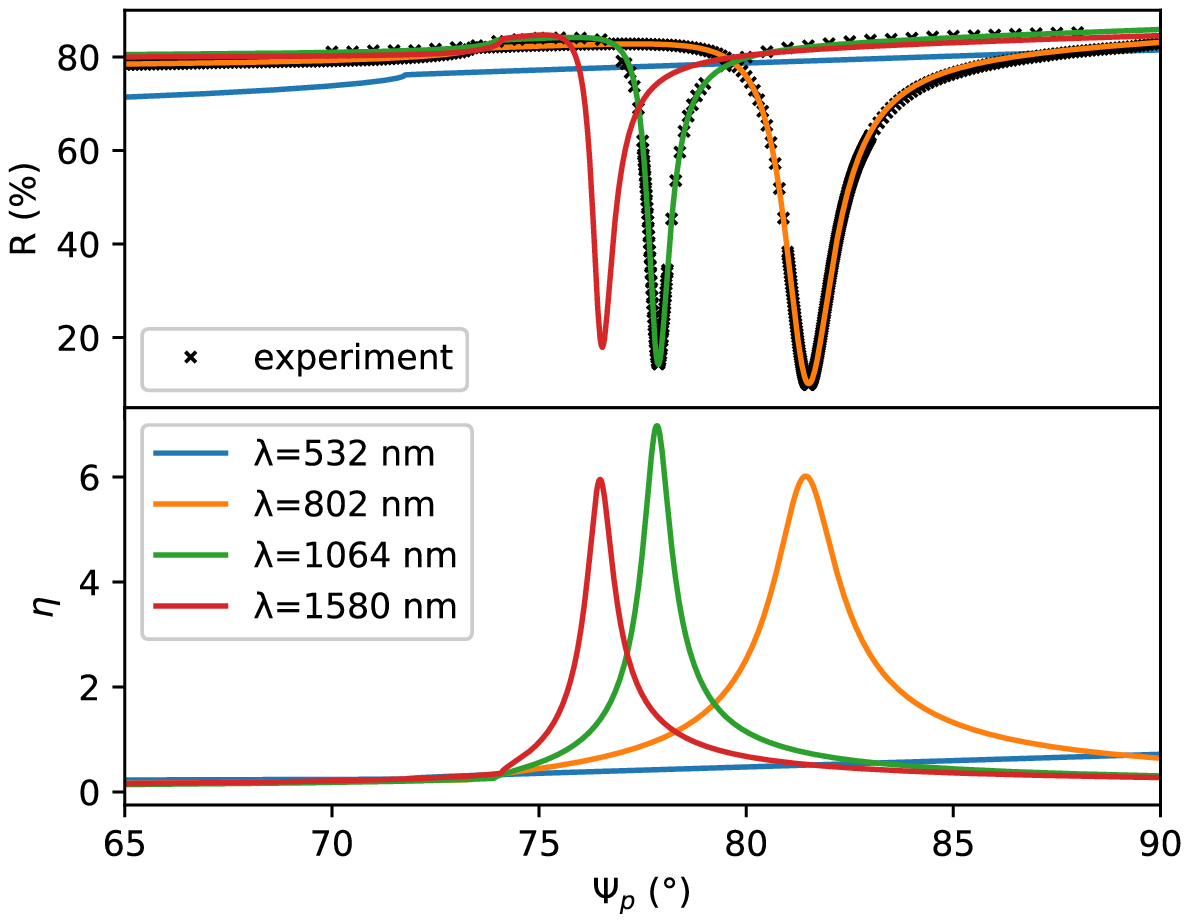} & \includegraphics[width=8cm]{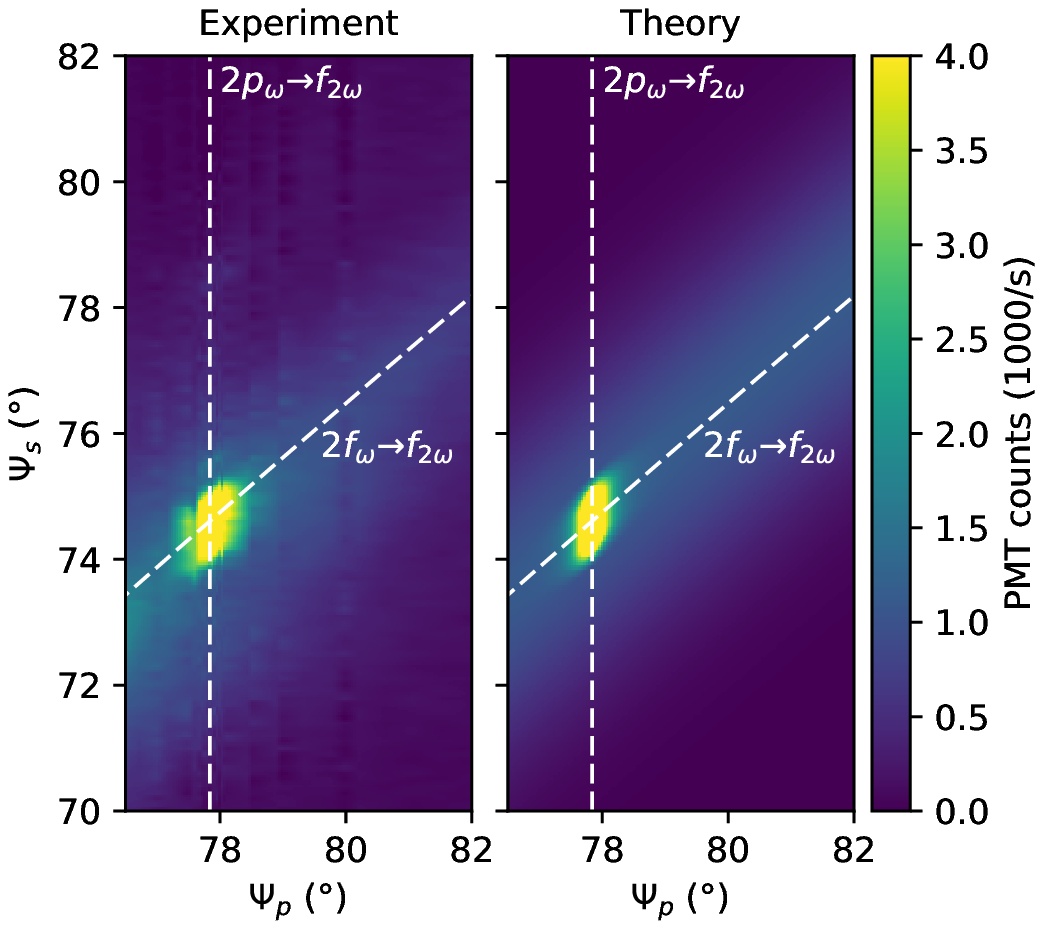}\tabularnewline
\end{tabular}
\par\end{centering}
\caption{a) Measured and modeled dependence of the reflection coefficient ($R$)
on the angle of incidence $\Psi_{p}$ of the pump beam along with
theoretical enhancement factor ($\eta$) . b) The experimentally measured
and theoretically predicted signal rates of the SPP-enhanced SHG.
\label{fig:spp-resonances-shg}}
\end{figure}

The measured reflection curves along with theoretical modeling are
shown in Fig. \ref{fig:spp-resonances-shg}a. Sharp plasmonic resonances
are clearly visible by a minimum in the reflection ($R$) and by a
maximum in the enhancement ($\eta$). The theoretical reflection/enhancement
curves were calculated by transfer-matrix method (TMM) and the unknown
parameters were fitted \citep{Abeles1957,Chilwell1984,Hodgkinson1997,Loot2015}.
Really good overlap of the experiment and the model was achieved and
the thicknesses and the complex refractive indices obtained from the
modeling are presented in Tab. \ref{tab:structure-fit-params}. The
refractive indices of the N-SF11 prism and the single crystal quartz
(ordinary component) perfectly match the known values from the literature
(www.schott.com and Ref. \citep{Ghosh1999a}). The Al\textsubscript{2}O\textsubscript{3}
layer in the model is effective and also accounts for the losses in
the IML. The fitting of the silver layer parameters revealed, that
indeed, the thickness of the layer is close to $50\,\text{nm}$ and
the complex refractive indices match quite well with the values reported
in Ref. \citep{Rakic1998}.

The reflection and the enhancement curves were also estimated for
wavelengths $532\,\text{nm}$ and $1580\,\text{nm}$ by using the
refractive index data from the literature \citep{Rakic1998,Ghosh1999a}.
The maximum field enhancement factors are up seven times (see Fig.
\ref{fig:spp-resonances-shg}a), which are considerably lower than
the values predicted in our previous theoretical calculations in Ref.
\citep{Loot2015}. This discrepancy is accounted for different refractive
index data for silver (data from P. B. Johnson and R. W. Christy \citep{Johnson1972a}),
which is known to significantly depend on the deposition and measurement
process \citep{Jiang2016}. Also, it shows, that the quality of the
silver film, used in this work, could be significantly improved by
the optimization of the deposition process.

\subsection{SPP-enhanced SHG}

Before studying SPP-enhanced SPDC, the structure was tested with simpler
SHG process. The enhancement of SHG by SPPs is well established both
theoretically and experimentally (see Refs. \citep{Second-harmonic1974,Simon1977,Kauranen2012,Grosse2012})
and thus can be used to confirm the nonlinearities of the structure. 

Here, we follow the methodology proposed by N. B. Grosse \citep{Grosse2012}
to study SPP-enhanced SHG by k-space spectroscopy \textendash{} we
scanned both the angle of incidence of the pump beam and the angle
of the sensor. Two different processes of SHG can take place in our
structure: two photons can produce another photon at twice the frequency
($2f_{\omega}\rightarrow f_{2\omega}$) and the pump laser can excite
two plasmons and subsequently produce one photon ($2p_{\omega}\rightarrow f_{2\omega}$).
The pure photonic process ($2f_{\omega}\rightarrow f_{2\omega}$)
can take place at any angle of incidence of the pump beam, however,
the plasmonic process ($2p_{\omega}\rightarrow f_{2\omega}$) can
only happen at the specific SPP resonance angle of the pump beam.

The experimental results are shown in Fig. \ref{fig:spp-resonances-shg}b
together with theoretical predictions. The weak SHG signal is clearly
visible along the predicted line of the process $2f_{\omega}\rightarrow f_{2\omega}.$
If the angle of incidence of the pump beam equals to the SPP-resonance
angle, then the process is greatly enhanced by plasmonic enhancement
(maximum in Fig. \ref{fig:spp-resonances-shg}b). It is direct evidence,
that our structure supports the plasmonic enhancement of nonlinear
processes. 

The theoretical enhancement in the right side of Fig. \ref{fig:spp-resonances-shg}b
is calculated by nonlinear TMM \citep{Loot2017}, normalized to experimentally
measured values and adjusted to the viewing angle of PMT. The match
between the experimentally measured and theoretically calculated values
is really good.

\subsection{SPP-enhanced SPDC}

\begin{figure}
\noindent \begin{centering}
\begin{tabular}{>{\raggedright}b{8cm}>{\raggedright}b{8cm}}
a) & b)\tabularnewline
\includegraphics[height=5.4cm]{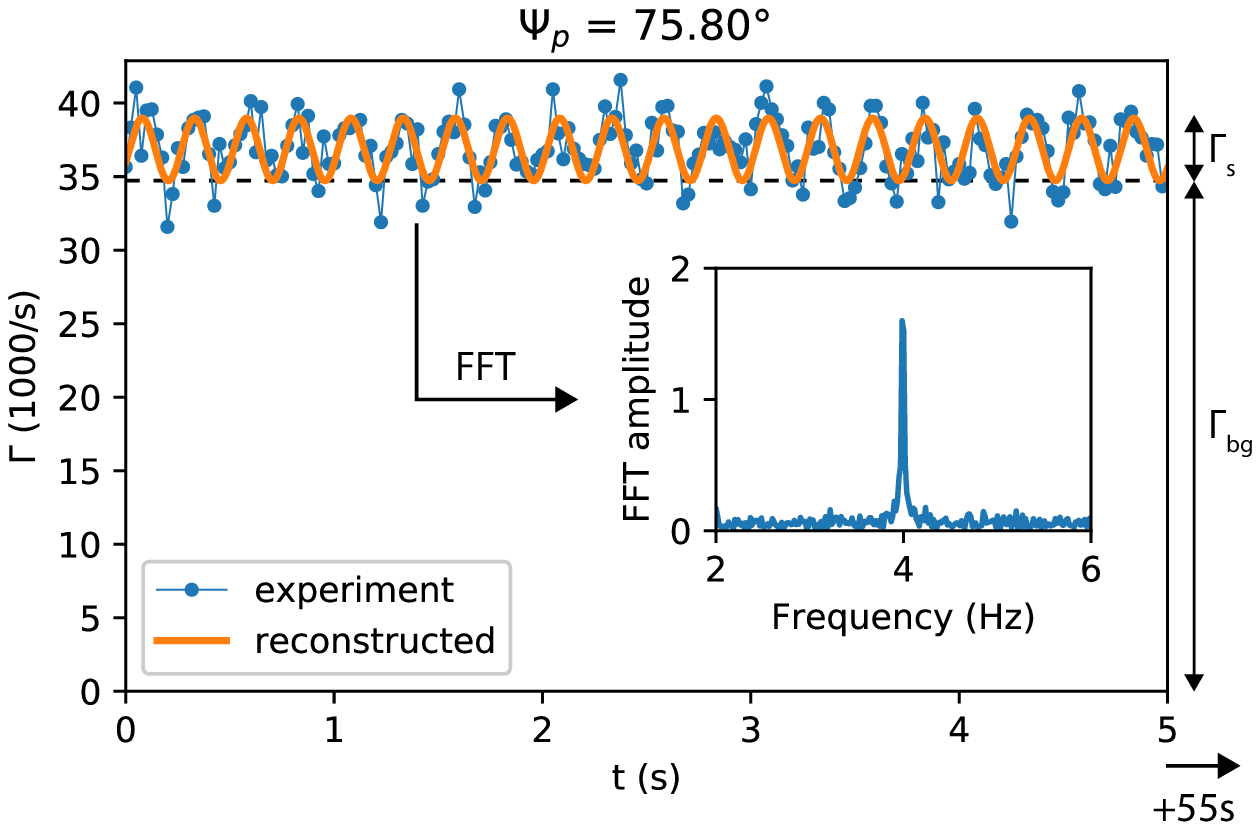} & \includegraphics[height=5.1cm]{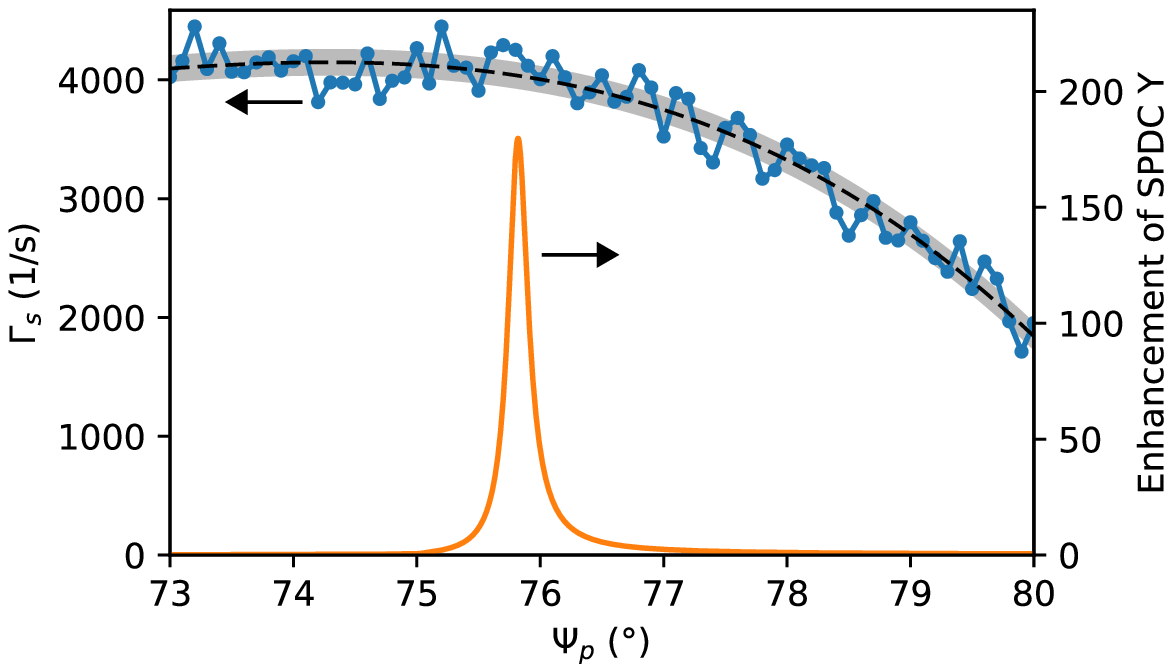}\tabularnewline
\end{tabular}
\par\end{centering}
\caption{a) An example of the detected time series at the angle of incidence
$\Psi_{p}=75.80^{\circ}$ along with reconstructed signal by lock-in
amplification. The Fourier spectrum of the detected signal is in the
inset. b) The dependence of the measured SPDC signal rates $\Gamma_{s}$
on the angle of incidence of the pump beam $\Psi_{p}.$ The theoretical
enhancement factor of the SPP-enhanced SPDC (Eq. \ref{eq:spdc-enh})
is given in the second axis.\label{fig:spdc-res-exp}}
\end{figure}

Our initial plan was to observe the SPP-enhanced SPDC by the $402\,\text{nm}$
laser and to detect the equally split signal and idler photons ($\lambda_{s}=\lambda_{i}=804\,\text{nm}$)
by PMT. However, incompatibility of the N-SF11 prism and IML with
the $402\,\text{nm}$ laser appeared \textendash{} the prism is luminescent
(see Fig. \ref{fig:experimental-setup-struct}d) and IML is not photostable
while irradiated by violet light and visibly changes its color from
transparent to brown in a couple of minutes.

As a workaround, the unequal splitting of $532\,\text{nm}\rightarrow800\,\text{nm}+1580\,\text{nm}$
(laser power $P=40\,\text{mW}$) was studied instead (like shown in
Fig. \ref{fig:spp-spdc-struct}). It is possible to select between
different splitting ratios by tuning the angle of incidence of the
pump beam $\Psi_{p}$ due to the phase-matching conditions (see Ref.
\citep{Loot2015}). 

Our strategy to detect the SPP-enhanced SPDC is the following. First,
only the photons at wavelength $800\,\text{nm}$ are detected due
to the sensitivity restriction of the PMT. Secondly, as the signal
is first generated into the plasmonic mode and then outcoupled to
the prism, we can easily experimentally determine the required sensor
angle. To do that, we use data from previous measurements of SPP resonance
at $\lambda=802\,\text{nm}$ (see Fig. \ref{fig:spp-resonances-shg}a):
the resonance happens at $\Psi_{p}=81.51^{\circ}$ and the correct
positioning of the sensor is at $\Psi_{s}=82.07^{\circ}.$ Note, that
the angle of incidence and the angle of the sensor are not the same
due to the refraction of light in the prism (discussed in Ref. \citep{Loot2013}).
Now, if the plasmons are generated at wavelength $800\,\text{nm},$
we know the correct positioning of the sensor in advance. Finally,
the only remaining free parameter is the angle of incidence of the
pump laser $\Psi_{p},$ which can be scanned to fulfill the phase-matching
conditions.

The detection of the signal by lock-in amplification is described
in Sec. \ref{subsec:Detection-of-SPDC} and is illustrated in Fig.
\ref{fig:spdc-res-exp}a for the angle of incidence $\Psi_{p}=75.80^{\circ}.$
Clear modulation of the signal at frequency $4\,\text{Hz}$ is visible
(Fourier spectrum is in the inset of Fig. \ref{fig:spdc-res-exp}a)
and the reconstructed signal clearly follows the noisy experimental
data.

The angle of incidence of the pump beam $\Psi_{p}$ was scanned over
range $73^{\circ}-80^{\circ}$ in steps of $0.1^{\circ}$ ($\approx1\,\text{hr}$
and $10\,\text{min}$) to achieve phase matching. It is expected to
have a single maximum near the angle $\Psi_{p}\approx75.8^{\circ}.$
The results are shown in Fig. \ref{fig:spdc-res-exp}b together with
the estimated enhancement factor of SPDC (Eq. \ref{eq:spdc-enh}).
However, no maximum near the expected angle is visible \textendash{}
the dependence of $\Gamma_{s}$ on the angle of incidence of the pump
beam is smooth and has a decreasing trend (black dashed line). The
gray shaded region in Fig. \ref{fig:spdc-res-exp}b represents the
theoretical limitation of the detection by shot noise ($95\,\%$ confidence
interval). The noise in the experimental data seems to be a bit higher,
but still close to the theoretical limitation.

From Fig. \ref{fig:spdc-res-exp}b it is estimated, that the noise
of the measurement is around $400\,\text{s}^{-1}.$ Taking into account
the power of the pump laser light and the quantum efficiency of the
PMT, the minimum detectable yield of SPDC is calculated to be around
$6\cdot10^{-14}$ \textendash{} the yield of SPP-enhanced SPDC must
be lower than it. 

\section{Discussion}

Here we discuss the three main possibilities why no SPP-enhanced SPDC
was detected. Firstly, the full modeling of SPP-enhanced SPDC in the
recent paper (Ref. \citep{Loot2018}) has revealed limiting factors
in addition to the plasmonic enhancement. The main limitation is the
short coherent buildup length ($\approx64\,\text{\ensuremath{\mu}m}$)
caused by the losses. However, the process of SPDC requires a long
coherent buildup distance (several millimeters) for the yield $10^{-12}$
and $64\,\text{\ensuremath{\mu}m}$ long coherent buildup distance
only permits yield $10^{-20}$ \textendash{} the yield is still small
($10^{-16}$) after the predicted enhancement by SPPs and below the
detection limit of our setup \citep{Loot2018}. Despite it, the SPP-enhanced
SPDC has still high potential in miniature sources: the short coherent
buildup is not a problem and the enhancement factor up to four orders
of magnitude is still available.

Secondly, only a fraction of the full enhancement of SPDC predicted
by a numerical modeling was experimentally available due to deficiencies
of the silver film \citep{Loot2015}. The fitting of reflection curves
(see Fig. \ref{fig:spp-resonances-shg}a) revealed, that the maximum
enhancement of SPPs is only around 7 \textendash{} several times lower
than predicted in Ref. \citep{Loot2015} on the basis of silver refractive
data from P. B. Johnson and R. W. Christy \citep{Johnson1972a}. The
optimization of the deposition of the silver film could further enhance
the SPDC by two orders of magnitude.

Thirdly, the lowest detectable yield of SPDC in our experimental setup
is limited to $6\cdot10^{-14}$ mainly by the high background signal
originating from the N-SF11 prism. To boost the sensitivity of the
experimental setup, the background rate must be reduced. To do that,
different prism or different excitation method (e.g. gratings) must
be used. As the prism must have a high refractive index ($>1.7$),
the selection of materials is quite limited.

\section{Conclusions}

In this communication, we have outlined a sensitive optical setup
and measurement protocol to experimentally realize the enhancement
of SPDC by SPPs. The details of the experimental setup were discussed
along with theoretical calculations of the enhancement factor. The
structure was characterized by the measurement of the reflection curves
and the nonlinearity of the structure was experimentally demonstrated
by SPP-enhanced SHG. However, no SPP-enhanced SPDC was detected mainly
due to the short coherent buildup distance in our structure.

\section{Acknowledgments }

This work was supported by the Estonian research projects IUT2-27,
IUT34-27 and the European Union through the European Regional Development
Fund (project 3.2.0101.11-0029 and Centre of Excellence 2014-2020.4.01.15-0011)
and through the ERA.Net Rusplus project \textquotedblleft 88 DIABASE\textquotedblright .

\section*{\textemdash \textemdash \textemdash \textemdash \textemdash \textendash{}}

\bibliographystyle{elsarticle-num-names}
\addcontentsline{toc}{section}{\refname}\bibliography{references}

\end{document}